# Glass transition and random packing in the hard sphere system

Hongqin Liu

CRCT, Ecole Polytechnique , University of Montreal

Box 6079, Station Downtown, Montreal, Quebec , CANADA, H3C 3A7

Email: hongqin.liu@polymtl.ca

**Glass transition and random packing in the hard sphere system have attracted great attention due to the important role of the system in the investigation of diverse real systems including liquids, colloidal dispersions, supercooled liquids, glasses and granular materials [1-12]. Despite the importance and "simplicity" of the system, some fundamental questions, such as the existence of an ideal glass transition [5-16], the nature of the glass transition and random packing (or jamming), and the entropy crisis or Kauzmann paradox [17,18], remain open. Based on a very accurate equation of state over the entire stable and metstable region within the potential energy landscape framework, here we report two "phase" transitions observed in the hard sphere system: the first is the ideal glass transition, indicating the configurational entropy vanishing, and the second, the jamming transition between the random loose packing and the random close packing [1,2] (or maximally random jammed packing [11,12]), indicating inherent structure domination. However, it is suggested that the glass and jamming transitions might not be treated as a thermodynamic phase transition. The unbalanced entropy loss suggests that equilibrium thermodynamics does not work for supercooled liquids and glasses. The results presented here for the hard sphere system will have direct impact on studies related to random packing or jamming and will shed a light on studies of glass transition in real systems.**

General discussions on glass transition and random packing (or jamming) can be found in the literature [3,4,17]. Here we focus on the hard sphere (HS) system, which is a simple model system with pure repulsive interaction. Glass transition in the HS system has been studied for decades [5-10]. The existence of a thermodynamic glass transition has been debated for decades since Gordon et al. suggested that HS glass transition is only a dynamic phenomenon [13]. While many authors are in favor of a thermodynamic phase transition or an ideal glass transition [5-10] (when configurational entropy vanishes), others found no evidence of such a transition [13-16]. No convincing signature has been found for an ideal glass transition.

A closely related issue is the entropy crisis, also known as the Kauzmann paradox [17,18]: because of the entropy loss in the transition process, the heat capacity of supercooled liquid/glass decreases as temperature decreases, and eventually it will become less than that of the crystal solid. This leads to a thermodynamic crisis: the entropy of the glass will become less than that of the crystal solid. For real systems this crisis can be avoided [3,19] and observed glass entropy at low temperature is very close to but not less than that of the crystal. The unbalanced entropy is assigned to that of the glass at zero temperature [3,19]. Will this thermodynamic balancing also work for the HS system? What can we learn from the entropy crisis (if any)?

Another very important issue is the random packing (jamming) in the HS system. In his well known experiment (1960) [1], Scott observed a random loose packing (RLP) and random close packing (RCP, recently defined as the maximally random jammed state, MRJ [11,12]). The subject has attracted considerable attention until recently [4,12,20], but the nature of random packing or jamming is still not completely understood.

For the glass transition analysis, the most useful thermodynamic quantity is the heat capacity, $C_P = (\partial H/\partial T)_P$, where $H$ is the enthalpy, $T$, the temperature, and $P$, the pressure. For real substances, $C_P$ can be measured directly with a calorimeter. For the HS system, since $H = \frac{3}{2}NkT + PV$, where $k$ is the Boltzmann constant, $N$, the number of particles, $V$, the total volume, the measurement of $C_P$ is equivalent to that of pressure, or compressibility, $Z = PV/NkT$. Therefore, a reliable analysis based on $C_P$ depends entirely on an accurate equation of state (EoS). Once a reliable EoS is given, $C_P$ can be calculated with the following equation [9]:

$$C_P = \frac{Z^2}{Z + \eta Z_\eta'} + \frac{3}{2} \quad (1)$$

where $\eta = \pi \rho^*/6$ is the packing fraction, $Z_\eta' = dZ/d\eta$. Other thermodynamic properties can also be calculated with the EoS, such as the excess entropy given by

$$s^{ex} = \frac{S - S^{ig}}{NK} = -\int_0^\eta \frac{Z-1}{\eta} d\eta \quad (2)$$

Excess entropy takes the ideal gas (ig) as reference and is mostly used in stable gaseous and liquid states. For supercooled liquid, glass and crystal solid, the absolute entropy is mostly employed and can be calculated by

$$s - s(0) = \frac{S - S(0)}{NK} = \int_0^T \frac{C_P}{T} dT \quad (3)$$

Here the reference state is at absolute zero temperature. For a perfect crystal solid, $s(0) = 0$, as required by the third law of thermodynamics. For a glass, it should be greater than zero. In the HS system, as seen below, this quantity diverges for both crystal and glass due to the finite values of heat capacity at zero temperature.

Because of lack of an accurate analytic EoS for the entire density range [21], stable and metstable regions are presently treated with two EoSs [9,10]: the Carnahan-Starling (CS) EoS [22] for the stable region and the Speedy EoS [7,8] for the metstable region, respectively. This is not only inconvenient, but also sometimes misleading as shown later. Here an EoS combining the features of potential energy landscape (PEL) approach and Woodcock EoS [5,6] is propsed (see Methods for details).

Tests based on a large number of computer simulation data demonstrate the high accuracy of the new EoS, Eq.(4), over the entire density range. In this EoS, the parameter, $\eta_{mrj}$, is of particular importance: it represents the MRJ state. The value obtained from the present work by the best fit of the compressibility data is 0.635591.

This value is consistent with those reported by different means: 0.6366 by the experiment of Scott & Kilgour [2], 0.64 obtained by computer simulation [11], and 0.63 by a more recent simulation [12]. This consistency and the high accuracy over the entire stable and metstable region guarantee that the new EoS, Eq.(4), will give reliable results for from zero density to the MRJ state.

Figure 1 depicts the inverse of the radial distribution function at contact, $g(\sigma) = (Z-1)/4\eta$, calculated with the new EoS and those from computer simulations. The results from the Carnahan-Starling EoS [22] for the stable region and the Speedy EoS [7,8] for the metstable region are also presented. The new EoS and computer simulation data clearly show that stable and metstable regions are smoothly connected branches, and a single EoS can be used to represent all four amorphous states of matter: gas, liquid, supercooled liquid and glass. The figure also indicates that in the neighborhood of density ~ 1.05, the CS EoS and the Speedy EoS cannot be applied and it is inappropriate to use these EoSs to probe the glass transition [9].

Figure 2a depicts the different components of the calculated compressibility in the PEL framework. It should be pointed out that the decomposition is not exact since the parameter fitting was performed when taking all the components together and the parameters in one component may be coupled with those in the other components. For an exact decomposition, independent inputs for two components are needed. This could not be accomplished due to lack of reliable data. Nevertheless, as shown below, the results given here do lead to reasonable conclusions.

Figure 2b illustrates the excess entropy calculated by Eq.(2) for supercooled liquid, glass and crystal solid. The Kauzmann paradox reported in reference [23] is also shown, which is a result of (equilibrium) extrapolation to high densities (low temperature) using the CS EoS. As shown in the figure, the excess entropy of supercooled liquids and glass calculated by the new EoS, Eq.(4), is always higher than that of the crystal solid, thus no thermodynamic law is violated.

Figure 3 presents the main findings of this work: phase transitions detected by heat capacity. Similar behavior was found using isothermal compressibility but omitted here for clarity. This figure contains a great deal of information to be discussed in detail.

First of all, some key quantities related to the glass transition [24] can now be accurately determined. The temperature at the peak is $T^*_{g,peak}$=0.071135 ($\rho^* = 1.05305$, $\eta = 0.55138$); at onset, $T^*_{g,onset}$=0.03454 ($\rho^* = 1.1478$, $\eta = 0.601$). There are two Kauzmann temperatures here. The first one is obtained by the extrapolation of "equilibrium" heat capacity. After references [10,23], the CS EoS was used for $C_p$ calculation, and by using the approach of Woodcock [6], the equilibrium Kauzmann temperature is found: $T^*_{K,eq}$=0.019 ($\rho^* = 1.183$, $\eta = 0.6194$), which is the same as that reported in reference [10]. (not surprisingly, this value is different from that obtained from excess entropy extrapolation, as shown in Figure 2, 0.0367). The second Kauzmann temperature is closely related to the ideal glass transition, indicating vanished configurational entropy. As shown in Figure 3a, it is the temperature at the intersection of the heat capacity ($C_P$) of the glass with that of the crystal solid: $T^*_K = 0.04938$ ($\rho^* = 1.11215$, $\eta = 0.58232$). Finally, the glass transition temperature, namely the inflection point of the slope of supercooled liquids/glass $C_p$, is given by Figure 3b: $T^*_g = 0.049048$ ($\rho^* = 1.113$, $\eta = 0.582766$), which is almost exact the same as $T^*_K$. Therefore, *hard evidence for an ideal glass transition is found*. By the way, the freezing point obtained by Hoover and Ree [25] is: $T^*_f = 0.12$ ($\rho^* = 0.9435$, $\eta = 0.494$, liquid phase). The ratio, $T^*_g/T^*_f \approx 0.41$.

Another very important result shown in Fig.3a and 3b is the second transition, which is similar to the first one. The $C_p$ peak appears at $T^*_{j,peak} = 0.016008$ ($\rho^* = 1.1905, \eta = 0.62334$), which is a precursor of a phase transition, and the transition temperature is given by Figure 3b: $T^*_J = 0.008368$ (*ρ\*=1.2075, η=0.63226*). This jamming transition appears between the random loose packing (RLP) and the random close packing (RCP, or MRJ state), first reported by Scott [1]. A more detailed discussion about the RLP and the RCP can be found in a recent publication [20] and references therein. The values of packing fraction at the RLP and the RCP can also be determined. For the RCP (or MRJ state): $\eta = 0.6356$; while for the RLP, the two possible choices are : (1) $T_{tr}^* = 0.026649$ ($\rho^*=1.1655$, $\eta=0.6102$) shown in Figure 3a, which is close to that ($\eta = 0.60$) observed by Scott [1]; or (2) $T_{RLP}^* = 0.0193$ ($\rho^*=1.1825$, $\eta=0.619$) shown in Figure 3b, and the same value has been reported by reference [20].

From Figure 3a and 3b, we can see that the jamming transition between the RLP and the RCP is also a "phase" transition, similar to a glass transition. This transition is likely a "poly-amorphous" glass transition in hard sphere system. Poly-amorphous transition between two glass states has been observed in some real systems, such as silicon [26]. For the HS system, are the transition between RLP and RCP and the "poly-amorphous" transition the same thing?

As for the ideal glass transition, its nature is now well understood: supercooled liquid loses its configurational entropy and becomes a glass. To understand why there is a second jamming transition, refer to Figure 2a. There is a key point in the figure: point A ($\eta=0.6326$), at which there is a structural change: IS contribution dominates the system: it starts to be jammed (particles trapped by their neighbors) until the MRJ point is reached. This point happens to be the jamming transition point ($\eta_J = 0.63226$). Therefore, a conjecture can be made: jamming transition is closely related to the IS domination when the vibrational contribution becomes less important (and can be ignored eventually). The jamming transition is related to the glass transition, but these two are different by nature.

A crisis can be immediately observed from Figure 3a: the entropy of glass is less than that of the crystal solid in the HS system. In the present application, the EoS is from "experimental" data fit, and heat capacity is calculated using the equation. The only thermodynamic relation involved is the enthalpy definition, and no entropy is involved. Therefore, the heat capacities shown in Figure 3 are reliable regardless equilibrium thermodynamics is applicable or not. For a quantitative analysis, some calculations have been carried out by using a thermodynamic approach discussed in the references [3, 19], in which unbalanced entropy is assigned to the glass entropy at zero temperature for real system. The details are given in Methods. The results unarguably show that the entropy loss during the glass transition can not be balanced for the HS system. Therefore, for the HS system, we inevitably reach the conclusion that *"thermodynamics does not work for glasses, because there is no equilibrium"* [27].

Now we try to answer the question: is glass transition a thermodynamic phase transition? From the above discussion, we see that equilibrium thermodynamics is not applicable to glass. Therefore, at least for the HS system, glass transition can not be considered as a conventional thermodynamic phase transition. Moreover, from Figure 3 we can see that both $C_P$ and $\partial C_P/\partial T$ show no discontinuity. In fact, the same is true for $\partial^2 C_P/\partial T^2$ and $d^3 C_P/dT^{*3}$. Therefore, it is unlikely that glass transition is of second or higher order. We suggest that the glass transition in the hard sphere system is not a thermodynamic phase transition.

For a complete picture of the HS system, we finally explore the glass transition with a property related to structural relaxation: the diffusion coefficient. Many accurate data have been reported up to high density [28-31]. Figure 4a plots the excess entropy scaling law [32] and the Arrhenius law. As seen from the figure, both laws break at temperature $T^* \approx 0.072$ ($\rho^* = 1.05$), just before the $C_P$ peak: ($T^*_{g,peak}$=0.071135, $\rho^* = 1.05305$), showing that the HS fluid becomes fragile [33] at this point.

Figure 4b depicts a scaled power law, $D^+ = (1-\eta/0.583)^{2.2}$. An excellent fitting is found by using the scaled property, $D^+ = D_{HS}[D_0 g(\sigma)]^{-1}$ where $D_0$ is the dilute gas diffusivity, suggested by Dzugutov [32]. The most interesting result from Figure 4b is that diffusivity of HS particles vanishes at packing fraction, *η=0.583*, which is exactly the glass transition point as shown in Figure 3b ($\eta_g = 0.582766$). Both Figure 4a and 4b suggest a strong connection between thermodynamics and structural relaxation in the glass transition.

## Methods

**Equation of state.** A significant achievement in the supercooled liquid and glass transition area is the potential energy landscape (PEL) approach [17]. In recent years, the PEL approach has been employed as a powerful tool for developing equation of state [34,35]. Within the PEL framework, the pressure of supercooled liquid/glass is expressed as [34,35] $P_{LT} = P_{IS} + P_{vib}$, where the subscript "LT" refers to low temperature. Then the total pressure can be written as: $P = P_{HT} + P_{IS} + P_{vib}$, where $P_{HT}$ is the contribution dominating the fluid behavior at high temperature; $P_{IS}$, the inherent structure (IS) contribution, which dominates the system structure at very low temperature and diverges as $T \to 0$; and $P_{vib}$, the vibrational contribution, which plays a significant role at low temperature but is less important at high temperature and does not diverge.

It is reasonable to adopt a truncated virial expansion [5,6] for high temperature contribution. For the IS contribution, the simple expression $a/(1-\alpha\eta)$ [6,8], which diverges at $\eta = \alpha^{-1}$, is a handy candidate. Finally, we need a function to account for the vibrational contribution. As mentioned above, this contribution is important in low temperature, while has a finite value as $T^* \to 0$. A natural, but somewhat arbitrary choice is a polynomial function with high power values. Combining all the three contributions, the new EoS proposed in this work reads

$$Z = 1 + \sum_{i=2}^{12} a_i \eta^{i-1} + \frac{c_0 \eta}{1 - \eta/\eta_{mrj}} + c_1 \eta^{27} + c_2 \eta^{29} + c_3 \eta^{31} \qquad (4)$$

The constants, $a_i$, are determined so that all exact virial coefficients up to the 10th are reproduced, the 11th and 12th are equal to the estimated values [36]. The values of constants $c_i$ ($i=0,…3$), the packing fraction at MRJ state, $\eta_{mrj}$, and the powers (27, 29, 31) are all adjusted parameters in order to obtain a minimum deviation between the compressibility values calculated with the EoS and those from computer simulations. A detailed discussion concerning the development of the EoS, its tests and a complete list of references for the data sources and numerical results are given in the Supplementary Materials. The correlation results are summarized as follows. For the stable region ($\rho^* = 0 \sim 0.95$) for 83 data points, Eq.(4) gives average absolute deviation (AAD) of only 0.12% (the CS EoS [22], 0.21%); for the metstable region ($\rho^* > 0.95$) for 95 data points, the AAD from Eq.(4) is 1.38% (the Speedy EoS [10,11], 5.82%, all other EoSs, give around or above 10%). The uncertainty of the new EoS is well within the simulation errors over the entire density range. The constants obtained are listed in Table 1.

**Table 1.** Values of constants of Eq.(4)

| $i$ | 1 | 2 | 3 | 4 | 5 | 6 | 7 |
|---|---|---|---|---|---|---|---|
| $a_i$ | 1 | 3.55173 | 9.29472 | 17.25512 | 26.47866 | 37.06833 | 49.02274 |
| $B_i$ | 1 | 4 | 10 | 18.36477 | 28.22451 | 39.81515 | 53.34442 |
| $i$ | 8 | 9 | 10 | 11 | 12 | | |
| $a_i$ | 61.7381 | 75.1150 | 88.9437 | 101.4448 | 111.008 | | |
| $B_i$ | 68.5376 | 85.8128 | 105.775 | 127.9263 | 152.6727 | | |

| $\alpha$ | $c_0$ | $c_1$ | $c_2$ | $c_3$ |
|---|---|---|---|---|
| 1.57334 | 0.44827 | $8.85739 \times 10^7$ | $-5.94896 \times 10^8$ | $1.0173 \times 10^9$ |

Notes: $\eta_{mrj} = 0.635591$; the relation between coefficient $a_i$ and virial coefficient $B_i$ is $B_i = a_i + c_0 \alpha^{i-1}$.

**Entropy balance**

The approach adopted here is the same as that used in references [3,19]. Taking an arbitrary point (at temperature $T$, "*" is dropped for clarity) on the $C_p$ curve of supercooled liquids/glass, there are two routes to compute the entropy by Eq.(3): the first route is on the liquid/glass curve,

$$s_{lg}(T) = s_{lg}(0) + \int_0^{T_K} C_{P,lg} d\ln T + \int_{T_K}^{T_f} C_{P,lg} d\ln T + \int_{T_f}^{T} C_{P,lg} d\ln T \qquad (5)$$

where the subscript "lg" stands for the supercooled liquids/glass, $T_K$ is the Kauzmann temperature, also the glass transition point, and $T_f$ is the freezing temperature of the liquid phase (or melting temperature of the crystal solid). The second route is on the crystal solid (subscript "cr") curve:

$$s_{lg}(T) = s_{cr}(0) + \int_0^{T_K} C_{P,cr} d\ln T + \int_{T_K}^{T_f} C_{P,cr} d\ln T - \Delta s_f + \int_{T_f}^{T} C_{P,lg} d\ln T \qquad (6)$$

where $\Delta s_f$ is the entropy of freezing. Combining Eq.(5) and Eq.(6), we have:

$$s_{cr}(0) - s_{lg}(0) + \int_0^{T_g} (C_{P,cr} - C_{P,lg}) d\ln T = \int_{T_g}^{T_f} (C_{P,lg} - C_{P,cr}) d\ln T + \Delta s_f \qquad (7)$$

Since in the region $0 \sim T_K$, $C_{P,cr} > C_{P,lg}$, as shown by Figure 3, the left hand side of Eq.(7) is always positive. The integration of the right hand side (RHS) of Eq.(7) can be carried out numerically using Eq.(4) for liquid/glass (present work) and $C_{P,cr}$ for crystal solid is from reference [6]. The result is: $\int_{T_g}^{T_f} (C_{P,lg} - C_{P,cr}) d\ln T \approx 0.29$. The value of $\Delta s_f$ has been evaluated for the HS system: -1.162 [25]. The sum of RHS of Eq.(7) is -0.872, a negative value! The entropy is not balanced using the above thermodynamic approach.

**Achnowledgement**

I thank Dr. Robin J Speedy, Dr. L. V. Woodcock, and Dr. S. Torquato for answering my questions regarding the related subjects and provide their articles. I also thank Dr. A. Pelton and Dr. S. Decterov (CRCT, Ecole Polytech.) for their support. Finally, I am grateful to my wife, Chongxia, for encouragement and support.

**References**

1. Scott, G. D. Packing of Spheres: Packing of Equal Spheres. Nature (London) 188, 908-909 (1960).

2. Scott, G. D. & Kilgour, D. M. The density of random close packing of spheres. Brit. J. Appl. Phys., ser.2 2, 863-866 (1969).

3. Debenedetti, P. G. *Metastable Liquids. Concepts and Principles* (Princeton Univ. Press, Princeton, 1996).

4. Torquato, S. *Random heterogeneous materials: microstructure and macroscopic properties* (Springer-Verlag, New York, 2002).

5. Woodcock, L. V. Glass transition in the hard - sphere model. J. Chem. Soc., Faraday Trans. 2: 72(9), 1667-1672 (1976).

6. Woodcock, L. V. Glass transition in the hard-sphere model and Kauzmann's paradox. Annals of the New York Academy of Sciences 371, 274-98 (1981).

7. Speedy, R. J. The hard sphere glass transition. Mol. Phys. 95, 169-178 (1998).

8 Speedy, Robin J. The equation of state for the hard sphere fluid at high density. The glass transition. Physica B+C: 121, 153-61 (1983).

9. Robles, M., Lopez H., M., Santos, A. & Bravo Yuste, S. Is there a glass transition for dense hard-sphere systems? J. Chem. Phys. 108, 1290-1291 (1998).

10. Parisi, G. & Zamponi, F. The ideal glass transition of hard spheres. J. Chem. Phys. 123, 144501/1-144501/12 (2005).

11. Torquato, S.; Truskett, T. M. & Debenedetti, P. G. Is Random Close Packing of Spheres Well Defined? Phys. Rev. Lett. 84, 2064-2067 (2000).

12. Kansal, A. R., Torquato, S. & Stillinger, F. H. Diversity of order and densities in jammed hard-particle packings. Phys. Rev. E 66, 41109/1-41109/8 (2002).

13. Gordon, J. M., Gibbs, J. H. & Fleming, P. D. The hard sphere "glass transition". J. Chem. Phys. 65, 2771-2778 (1976).

14. Rintoul, M. D. & Torquato, S. Metastability and crystallization in hard-sphere systems. Phys. Rev. Lett. 77, 4198-4201 (1996).

15. Torquato, S. Glass transition: Hard knock for thermodynamics. Nature (London) 405, 521-523 (2000).

16. Donev A., F. H. Stillinger, F. H. & S. Torquato, Do Binary Hard Disks Exhibit an Ideal Glass Transition? arXiv.org, cond-mat/0603183/1-4

17. Debenedetti, P. G. & Stillinger, F. H. Supercooled liquids and glass transition. Nature, 410, 259-267 (2001).

18. Kauzmann, W. The nature of the glassy state and the behavior of liquids at low temperatures. *Chem. Rev.* **43,** 219–256 (1948).

19. Johari, G. P.. The configurational entropy theory and the heat capacity decrease of orientationally disordered crystals on cooling to 0 K. Philosophical Magazine B: 81, 1935-1950 (2001).

20 Dong, K. J., Yang, R. Y., Zou, R. P. & Yu, A. B. Role of interparticle forces in the formation of random loose packing. Phys. Rev. Lett. 96, 145505/1-145505/5 (2006).

21. Kolafa, J. Nonanalytical equation of state of the hard sphere fluid. Phys. Chem. Chem. Phys. 8, 464-468 (2006).

22. Carnahan, N. F. & Starling, K. E.. Equation of state for nonattracting rigid spheres. J. Chem. Phys. 51, 635-636 (1969).

23. Stillinger, F. H.; Debenedetti, P. G. & Truskett, T. M. The Kauzmann Paradox Revisited. J. Phys. Chem. B 105, 11809-11816 (2001).

24. Angell, C. A. Liquid fragility and the glass transition in water and aqueous solutions. Chem. Rev. 102, 2627-2650 (2002).

25. Hoover, W. G. & Ree, F. H. Melting Transition and Communal Entropy for Hard Spheres . J. Chem. Phys. 49, 3609-3618 (1968).

26. Hedler, A., Klaumunzer, S. L. & Wesch, W. Amorphous silicon exhibits a glass transition. Nature Materials, 3, 804-809 (2004).

27. Nieuwenhuizen, Th M. Thermodynamic picture of the glassy state. J. Physics: Condensed Matter 12, 6543-6552 (2000).

28. Woodcock, L. V. & Angell, C. A. Diffusivity of the Hard-Sphere Model in the Region of Fluid Metastability**.** Phys. Rev. Lett. 47, 1129-1132 (1981).

29. Erpenbeck, J. J. & Wood, W. W. Self- diffusion coefficient for the hard-sphere fluid. Phys. Rev. A 43, 4254-61 (1991).

30. Sigurgeirsson, H.; Heyes, D. M. Transport coefficients of hard sphere fluids. Mol. Phys. 101, 469-482 (2003).

31. Kumar, S. K, Szamel, G. & Douglas, J. F. Nature of the breakdown in the Stokes-Einstein relationship in a hard sphere fluid. arXiv.org, cond-mat/0508172

32. Dzugutov, M. A universal scaling law for atomic diffusion in condensed matter. Nature 381, 137-139 (1996).

33. Angell, C. A. Formation of glasses from liquids and biopolymers. Science 267, 1924-1935 (1995).

34. Debenedetti, P. G., Stillinger, F. H., Truskett, T. M. & Roberts, C. J. The equation of state of an energy landscape. *J. Phys. Chem. B* **103,** 7390–7397 (1999).

35. Shell, M. S., Debenedetti, P. G., La Nave, E. & Sciotino, F. Energy landscapes, ideal glasses, and their equation of state. J. Chem. Phys. 118, 8821-8830 (2003).

36. Clisby, N. & McCoy, B. M. Ninth and tenth order virial coefficients for hard spheres in D dimensions. arXiv:cond-mat/0503525. 1-34 (2005).

**Figure captions**

**Figure 1**. Plot of $1/g(\sigma) \sim \eta$.

**a.** The entire density region.
Points are computer simulation data. Solid line, Eq.(4). The inner panel illustrates the metstable region detailed in Figure 1b. The figure also shows that stable and metstable regions are smoothly connected.

**b.** The metstable region.
Points are computer simulation data. Solid line, Eq.(4); dashed line, the CS EoS [22]; dotted line, the Speedy EoS [8]. Notice that both the CS EoS and the Speedy EoS start to leave the right track in the neighborhood of $\eta = 0.55$, where heat capacity peaks.

**Figure 2** The decomposed compressibility plot and the excess entropy plot.

**a**. The decomposed compressibility.
The IS contribution is calculated by $Z_{IS} = 0.2947 \left[ 1/(1-1.57334\eta) - \sum_{i=0}^{11} (1.57334\eta)^i \right]$.
Point A ($\eta=0.6326$) is where IS starts to dominate the system structure.

**b**. The excess entropy.
Solid line: supercooled liquids/glass, Eq.(4); dotted line, "equilibrium" liquid, the CS EoS [22]; dashed line, the crystal solid (Woodcock [6]).

**Figure 3**. Heat capacity and its first derivative.

**a**. Heat capacity plot. The superscript "*" was dropped in the figure for clarity. Some key quantities can be determined by this Figure. $T^*_{g,peak} = 0.071135$ ($\rho^*=1.05305$, $\eta=0.55138$), $T^*_K = 0.04938$, the intersection temperature of $C_{p,\text{lg}}$ and $C_{p,cr}$, ($\rho^*=1.11215$, $\eta=0.58232$), $T^*_{g,onset} = 0.03454$ ($\rho^*=1.1478$, $\eta=0.601$), $T^*_{tr} = 0.026649$, the transit point from the first peak to the second peak, ($\rho^*=1.1655$, $\eta=0.610255$), and the second peak, $T^*_{j,peak} = 0.016008$ ($\rho^*$=1.1905, $\eta$=0.62334). The Kauzmann temperature by $C_{p,eq}$ estimated from CS EoS is $T^*_K = 0.019094$ ($\rho^*$=1.183, $\eta$=0.6194), not shown.

**b.** The first derivative of heat capacity, $\partial C_P/\partial T$. Three key quantities can be determined. The glass transition temperature, $T^*_g = 0.049048$ ($\rho^*$=1.113, $\eta$=0.582766), which is very close to $T^*_K = 0.04938$ ($\eta$=0.58232). The jamming transition temperature $T^*_J = 0.008368$ ($\rho^*$=1.2075, $\eta$=0.63225). The valley point can be defined as the random loose packing point: $T^*_{RLP} = 0.0193$ ($\rho^*$=1.1825, $\eta$=0.619).

**Figure 4**. Diffusion coefficient plot.

**a**. Arrhenius law (circles) and excess entropy scaling law (triangles) plots. The lines are guides to the eye. Both laws break at $T^* = 0.0714$ ($T^*_{g,peak} = 0.071135$, from $C_P$ plot). The constant -2.5 was introduced for convenience of the plotting. The scaled diffusivity is defined as $D^+ = D_{HS}[D_0 g(\sigma)]^{-1}$, where $D_0$ is the diffusivity of dilute gas. If the quantity $D^* = D_{HS}/D_0$ is used instead of $D^+$, similar results can be obtained.

**b**. Scaled power law plot. The points are simulation data. Line is the power law: $D^+ = (1-\eta/0.583)^{2.2}$.

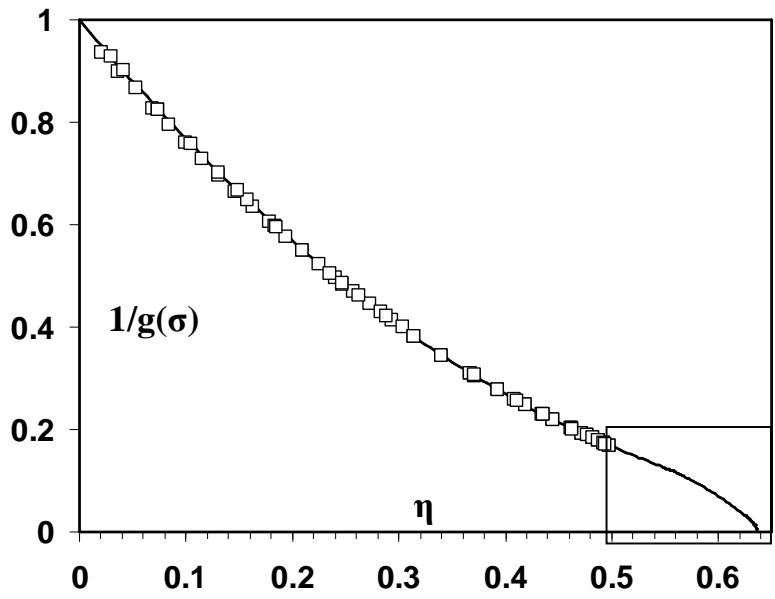

1a

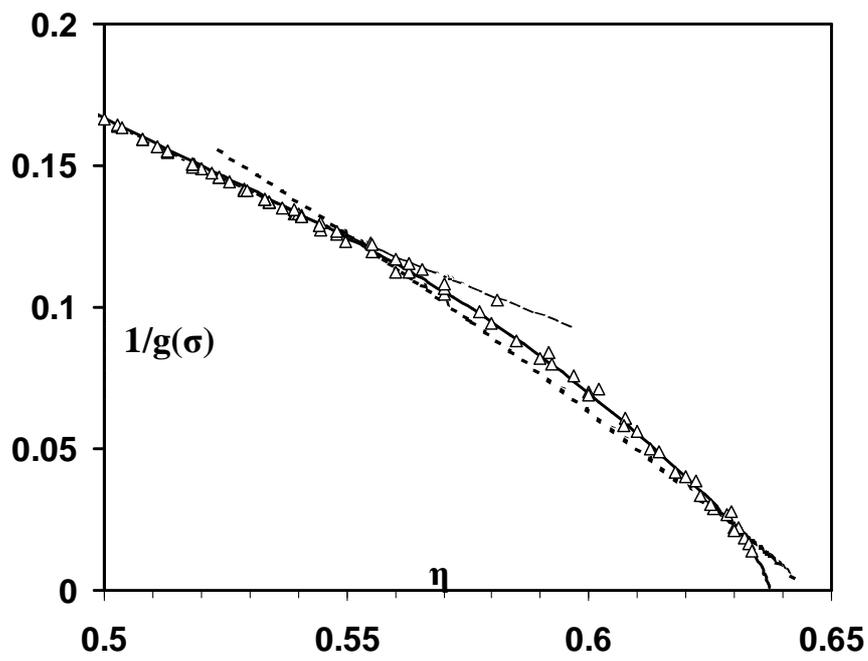

1b

Figure 1.

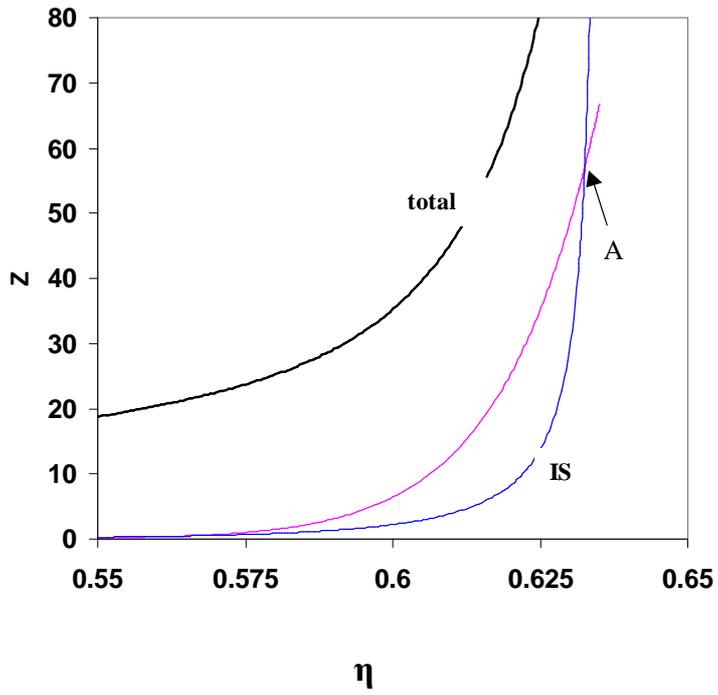

2a.

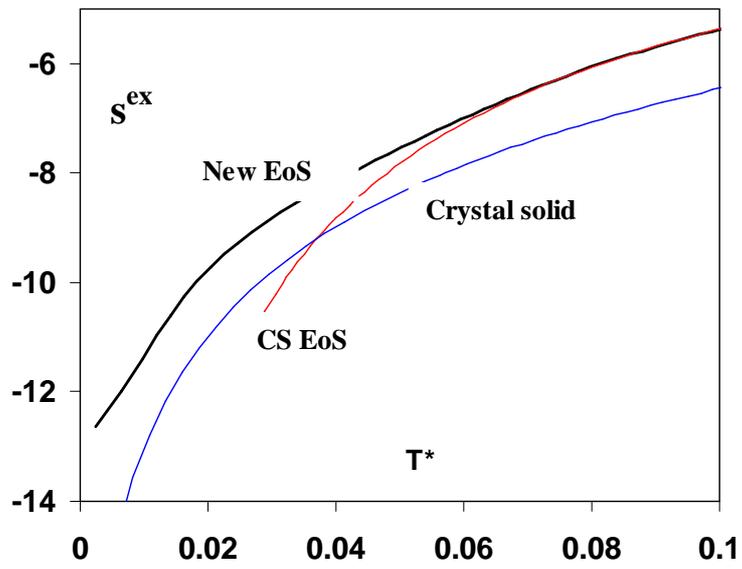

2b.

Figure 2

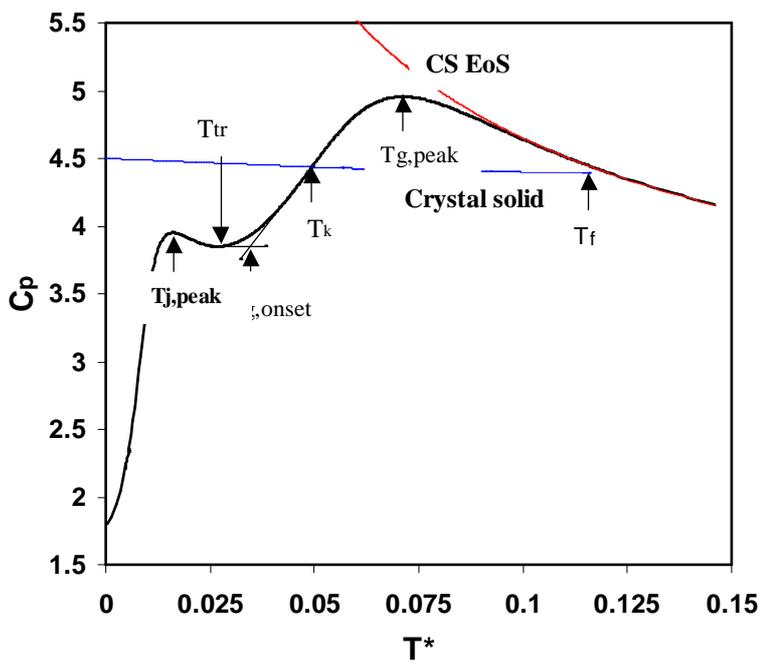

3a

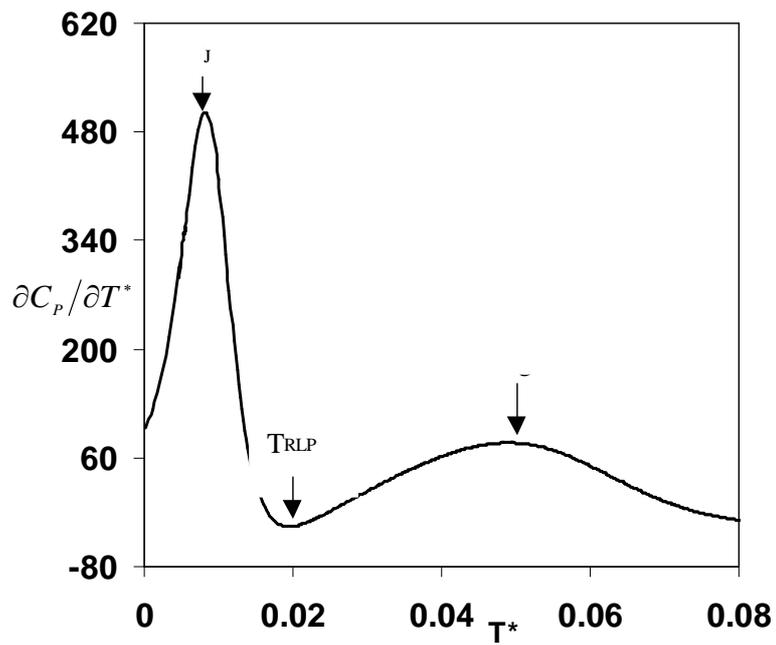

3b

Figure 3

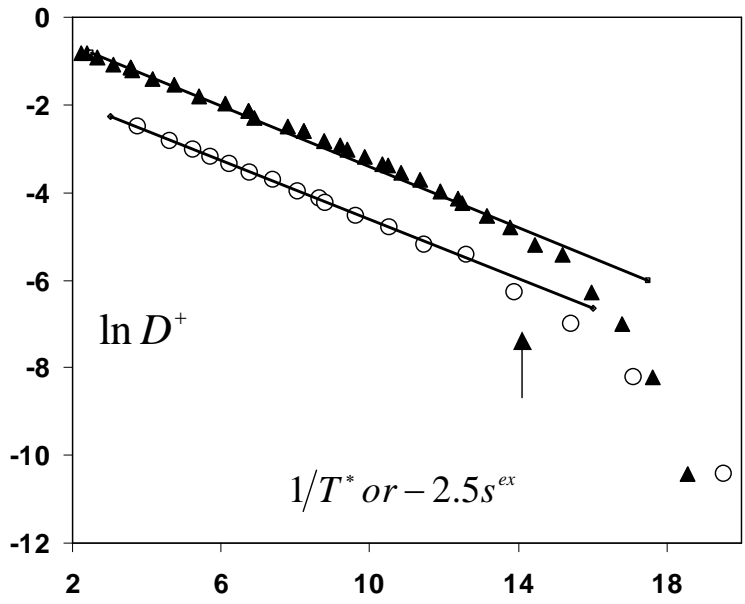

4a

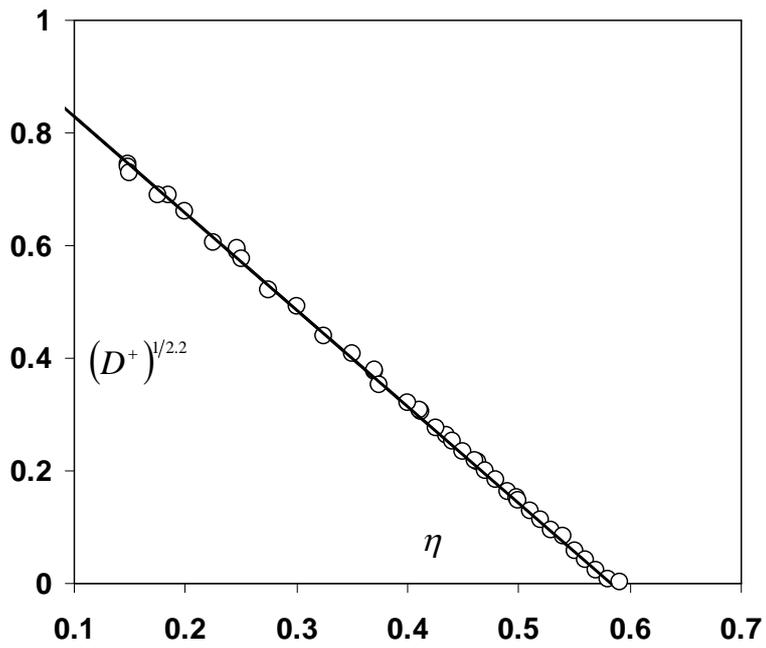

4b

Figure 4